\documentclass[a4paper]{report}
\usepackage[utf8]{inputenc}
\usepackage[T1]{fontenc}
\usepackage{RJournal}
\usepackage{amsmath,amssymb,array}
\usepackage{booktabs}

\providecommand{\tightlist}{%
  \setlength{\itemsep}{0pt}\setlength{\parskip}{0pt}}

\usepackage{float}

\begin{document}

\sectionhead{Contributed research article}
\volume{XX}
\volnumber{YY}
\year{20ZZ}
\month{AAAA}

\begin{article}
\title{brolgar: An R package to BRowse Over Longitudinal Data
Graphically and Analytically in R}
\author{by Nicholas Tierney, Dianne Cook, and Tania Prvan}

\maketitle

\abstract{%
Longitudinal (panel) data provide the opportunity to examine temporal
patterns of individuals, because measurements are collected on the same
person at different, and often irregular, time points. The data is
typically visualised using a ``spaghetti plot'', where a line plot is
drawn for each individual. When overlaid in one plot, it can have the
appearance of a bowl of spaghetti. With even a small number of subjects,
these plots are too overloaded to be read easily. The interesting
aspects of individual differences are lost in the noise. Longitudinal
data is often modelled with a hierarchical linear model to capture the
overall trends, and variation among individuals, while accounting for
various levels of dependence. However, these models can be difficult to
fit, and can miss unusual individual patterns. Better visual tools can
help to diagnose longitudinal models, and better capture the individual
experiences. This paper introduces the R package, brolgar (BRowse over
Longitudinal data Graphically and Analytically in R), which provides
tools to identify and summarise interesting individual patterns in
longitudinal data.
}

\hypertarget{introduction}{%
\section{Introduction}\label{introduction}}

This paper is about exploring longitudinal data effectively.
Longitudinal data can be defined as individuals repeatedly measured
through time, and its inherent structure allows us to examine temporal
patterns of individuals. This structure is shown in Figure
\ref{fig:heights-sample-plot}, which shows a sample of data from the
average height of Australian males. The individual component is
\texttt{country}, and the time component is \texttt{year}. The variable
\texttt{country} along with other variables is measured repeatedly from
1900 to 1970, with irregular time periods between years.

\begin{Schunk}
\begin{figure}

{\centering \includegraphics[width=0.95\linewidth]{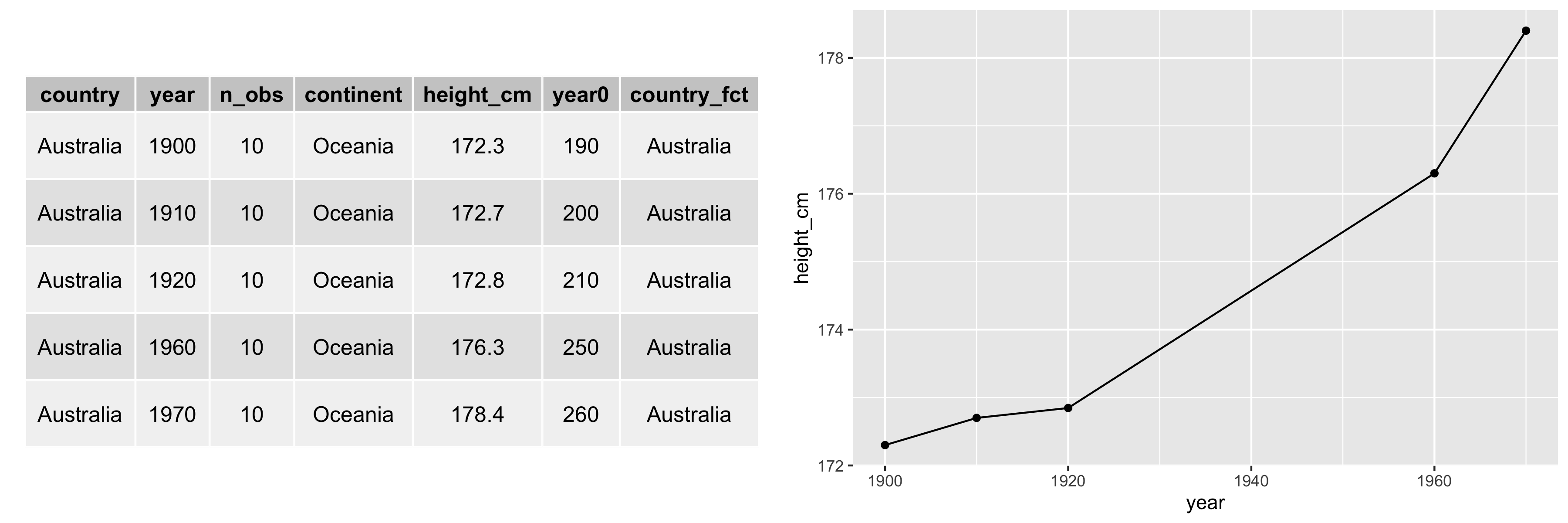} 

}

\caption[A sample of average heights of men in Australia for 1900-1970]{A sample of average heights of men in Australia for 1900-1970. The heights increase over time, but are measured at irregular intervals.}\label{fig:heights-sample-plot}
\end{figure}
\end{Schunk}

The full dataset of Figure \ref{fig:heights-sample-plot} is shown in
Figure \ref{fig:spaghetti}, showing 144 countries from the year 1700.
This plot is challenging to understand because there is overplotting,
making it hard to see the individuals. Solutions to this are not always
obvious. Showing separate individual plots of each country does not
help, as 144 plots is too many to comprehend. Making the lines
transparent or fitting a simple model to all the data Figure
\ref{fig:spaghetti}B, might be a common first step to see common trends.
However, all this seems to clarify is: 1) There is a set of some
countries that are similar, and they are distributed around the center
of the countries, and 2) there is a general upward trend in heights over
time. We learn about the collective, but lose sight of the individuals.

\begin{Schunk}
\begin{figure}

{\centering \includegraphics[width=0.95\linewidth]{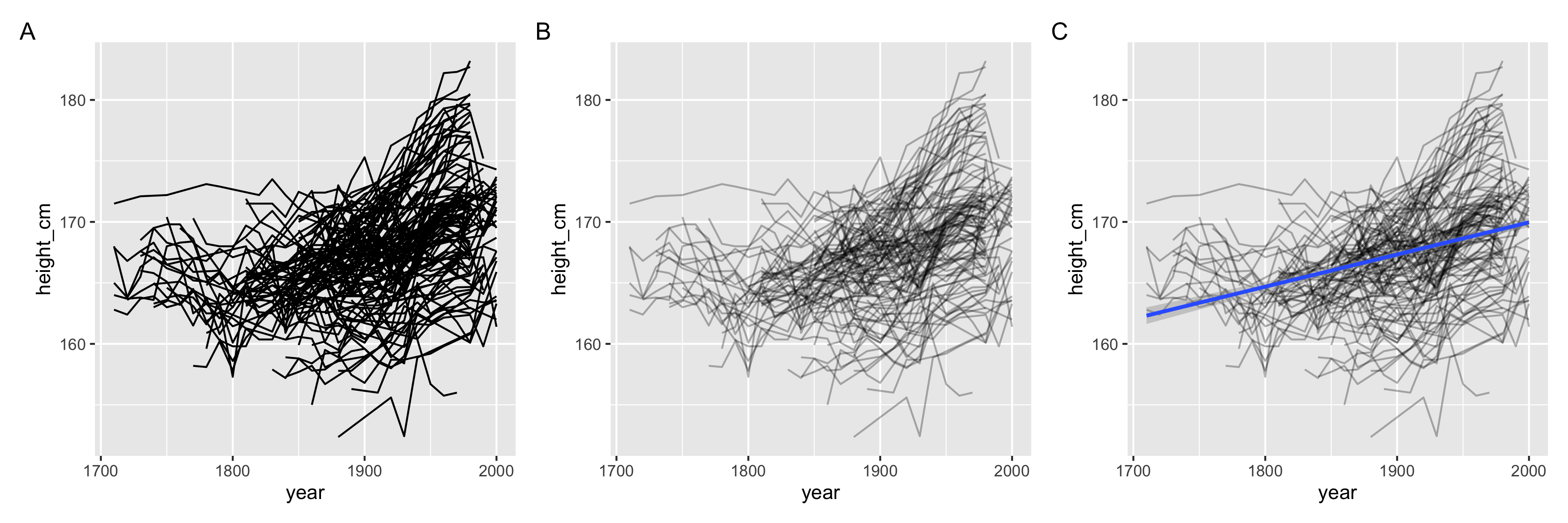} 

}

\caption[The full dataset shown as a spaghetti plot (A), with transparency (B), and with a linear model overlayed (C)]{The full dataset shown as a spaghetti plot (A), with transparency (B), and with a linear model overlayed (C). It is still hard to see the individuals.}\label{fig:spaghetti}
\end{figure}
\end{Schunk}

This paper demonstrates how to effectively and efficiently explore
longitudinal data, using the R package, \texttt{brolgar}. We examine
four problems in exploring longitudinal data:

\begin{enumerate}
\def\labelenumi{\arabic{enumi}.}
\tightlist
\item
  How to sample the data
\item
  Finding interesting individuals
\item
  Finding representative individuals
\item
  Understanding a model
\end{enumerate}

This paper proceeds in the following way: first, a brief review of
existing approaches to longitudinal data, then the definition of
longitudinal data, then approaches to these four problems are discussed,
followed by a summary.

\hypertarget{other-approaches}{%
\section{Background}\label{other-approaches}}

R provides basic time series, \texttt{ts}, objects, which are vectors or
matrices that represent data sampled at equally spaced points in time.
These have been extended through packages such as \texttt{xts}, and
\texttt{zoo} \citep{xts, zoo}, which only consider data in a wide format
with a regular implied time series. These are not appropriate for
longitudinal data, which can have indexes that are not time unit
oriented, such as ``Wave 1\ldots n'', or may contain irregular
intervals.

Other packages focus more directly on panel data in R, focussing on data
operations and model interfaces. The \texttt{pmdplyr} package provides
``Panel Manoeuvres'' in ``dplyr'' \citep{pmdplyr}. It defines the data
structure in as a \texttt{pibble} object (\textbf{p}anel
t\textbf{ibble}), requiring an \texttt{id} and \texttt{group} column
being defined to identify the unique identifier and grouping. The
\texttt{pmdplyr} package focuses on efficient and custom joins and
functions, such as \texttt{inexact\_left\_join()}. It does not implement
tidyverse equivalent tools, but instead extends their usecase with a new
function, for example \texttt{mutate\_cascade} and
\texttt{mutate\_subset}. The \texttt{panelr} package provides an
interface for data reshaping on panel data, providing widening and
lengthening functions (\texttt{widen\_panel()} and
\texttt{long\_panel()} \citep{panelr}). It also provides model
facilitating functions by providing its own interface for mixed effects
models.

These software generally re-implement their own custom panel data class
object, as well as custom data cleaning tasks, such as reshaping into
long and wide form. They all share similar features, providing some
identifying or index variable, and some grouping or key.

\hypertarget{data}{%
\section{Longitudinal Data Structures}\label{data}}

Longitudinal data is known by many names: Panel data, survey data,
repeated measures, and time series, to name a few. Although there are
small differences among these definitions related to data collection,
context, and field, these data structures all share a fundamental
similarity: they are measurements of the same individual over a time
period.

This time period has structure - the time component (dates, times,
waves, seconds, etc), and the spacing between measurements - unequal or
equal. This data structure needs to be respected during analysis to
preserve the lowest level of granularity, to avoid for example,
collapsing across month when the data is collected every second, or
assuming measurements occur at fixed time intervals. These mistakes can
be avoided by encoding the data structure into the data itself. This
information can then be accessed by analysis tools, providing a
consistent way to understand and summarise the data. This ensures the
different types of longitudinal data previously mentioned can be handled
in the same way.

\hypertarget{building-on-tsibble}{%
\subsection{\texorpdfstring{Building on
\texttt{tsibble}}{Building on tsibble}}\label{building-on-tsibble}}

Since longitudinal data can be thought of as ``individuals repeatedly
measured through time'', they can be considered as a type of time
series, as defined in \citet{fpp}: ``Anything that is observed
sequentially over time \textbf{is a time series}''. This definition has
been realised as a time series \texttt{tsibble} in \citep{Wang2020}.
These objects are defined as data meeting these conditions:

\begin{enumerate}
\def\labelenumi{\arabic{enumi}.}
\tightlist
\item
  The \texttt{index}: the time variable
\item
  The \texttt{key}: variable(s) defining individual groups (or series)
\item
  The \texttt{index} and \texttt{key} (1 + 2) together determine a
  distinct row
\end{enumerate}

If the specified key and index pair do not define a distinct row - for
example, if there are duplicates in the data, the \texttt{tsibble} will
not be created. This helps ensure the data is properly understood and
cleaned before analysis is conducted, removing avoidable errors that
might have impacted downstream decisions.

We can formally define our \texttt{heights} data from Figure
\ref{fig:heights-sample-plot} as a \texttt{tsibble} using,
\texttt{as\_tsibble}:

\begin{Schunk}
\begin{Sinput}
heights_brolgar <- as_tsibble(heights_brolgar,
                      index = year,
                      key = country,
                      regular = FALSE)
\end{Sinput}
\end{Schunk}

The \texttt{index} is \texttt{year}, the \texttt{key} is
\texttt{country}, and \texttt{regular\ =\ FALSE} since the intervals in
the years measured are not regular. Using a \texttt{tsibble} means that
the index and key time series information is recorded only
\textbf{once}, and can be referred to many times in other parts of the
data analysis by time-aware tools.

In addition to providing consistent ways to manipulate time series data,
further benefits to using \texttt{tsibble} are how it works within the
\texttt{tidyverse} ecosystem, as well as the tidy time series packages
called ``tidyver\textbf{ts}'', containing \texttt{fable} \citep{fable},
\texttt{feasts}, \citep{feasts}. For example, \texttt{tsibble} provides
\texttt{tidyverse} extension functions to explore implicit missing
values in the \texttt{index} (e.g., \texttt{has\_gaps()} and
\texttt{fill\_gaps()}), as well as grouping and partitioning based on
the index with \texttt{index\_by()}. For full details and examples of
use with the tidyverts time series packages, see \citet{Wang2020}.

The \texttt{brolgar} package uses \texttt{tsibble} so users can take
advantage of these tools, learning one way of operating a data analysis
that will work and have overlap with other contexts.

\hypertarget{characterising-individual-series}{%
\subsection{Characterising Individual
Series}\label{characterising-individual-series}}

\hypertarget{calculating-a-feature}{%
\subsubsection{Calculating a Feature}\label{calculating-a-feature}}

We can summarise the individual series by collapsing their many
measurements into a single statistic, such as the minimum, maximum, or
median, with one row per key. We do this with the \texttt{features}
function from the \texttt{fabletools} package, made available in
\texttt{brolgar}. This provides a summary of a given variable,
accounting for the time series structure, and returning one row per key
specified. It can be thought of as a time-series aware variant of the
\texttt{summarise} function from \texttt{dplyr}. The \texttt{feature}
function works by specifying the data, the variable to summarise, and
the feature to calculate. A template is shown below

\begin{verbatim}
features(<DATA>, <VARIABLE>, <FEATURE>)
\end{verbatim}

or, with the pipe:

\begin{verbatim}
<DATA> %>% features(<VARIABLE>, <FEATURE>)
\end{verbatim}

For example, to calculate the minimum height for each key (country), in
\texttt{heights}, we specify the \texttt{heights} data, then the
variable to calculate features on, \texttt{height\_cm}, then the feature
to calculate, \texttt{min} (we write \texttt{c(min\ =\ min)} so the
column calculated gets the name ``min''):

\begin{Schunk}
\begin{Sinput}
heights_min <- features(.tbl = heights_brolgar, 
                        .var = height_cm, 
                        features = c(min = min))

heights_min
\end{Sinput}
\begin{Soutput}
#> # A tibble: 119 x 2
#>    country       min
#>    <chr>       <dbl>
#>  1 Afghanistan  161.
#>  2 Algeria      166.
#>  3 Angola       159.
#>  4 Argentina    167.
#>  5 Armenia      164.
#>  6 Australia    170 
#>  7 Austria      162.
#>  8 Azerbaijan   170.
#>  9 Bangladesh   160.
#> 10 Belgium      163.
#> # ... with 109 more rows
\end{Soutput}
\end{Schunk}

We call these summaries \texttt{features} of the data. We can use this
information to summarise these features of the data, for example,
visualising the distribution of minimum values (Figure
\ref{fig:feature-min-med-max}A)

We are not limited to one feature at a time, many features can also be
calculated, for example:

\begin{Schunk}
\begin{Sinput}
heights_three <- heights_brolgar %>%
  features(height_cm, c(
    min = min,
    median = median,
    max = max
  ))

heights_three
\end{Sinput}
\begin{Soutput}
#> # A tibble: 119 x 4
#>    country       min median   max
#>    <chr>       <dbl>  <dbl> <dbl>
#>  1 Afghanistan  161.   167.  168.
#>  2 Algeria      166.   169   171.
#>  3 Angola       159.   167.  169.
#>  4 Argentina    167.   168.  174.
#>  5 Armenia      164.   169.  172.
#>  6 Australia    170    172.  178.
#>  7 Austria      162.   167.  179.
#>  8 Azerbaijan   170.   172.  172.
#>  9 Bangladesh   160.   162.  164.
#> 10 Belgium      163.   166.  177.
#> # ... with 109 more rows
\end{Soutput}
\end{Schunk}

These can then be visualised together (Figure
\ref{fig:feature-min-med-max}).

\begin{Schunk}
\begin{figure}

{\centering \includegraphics[width=0.95\linewidth]{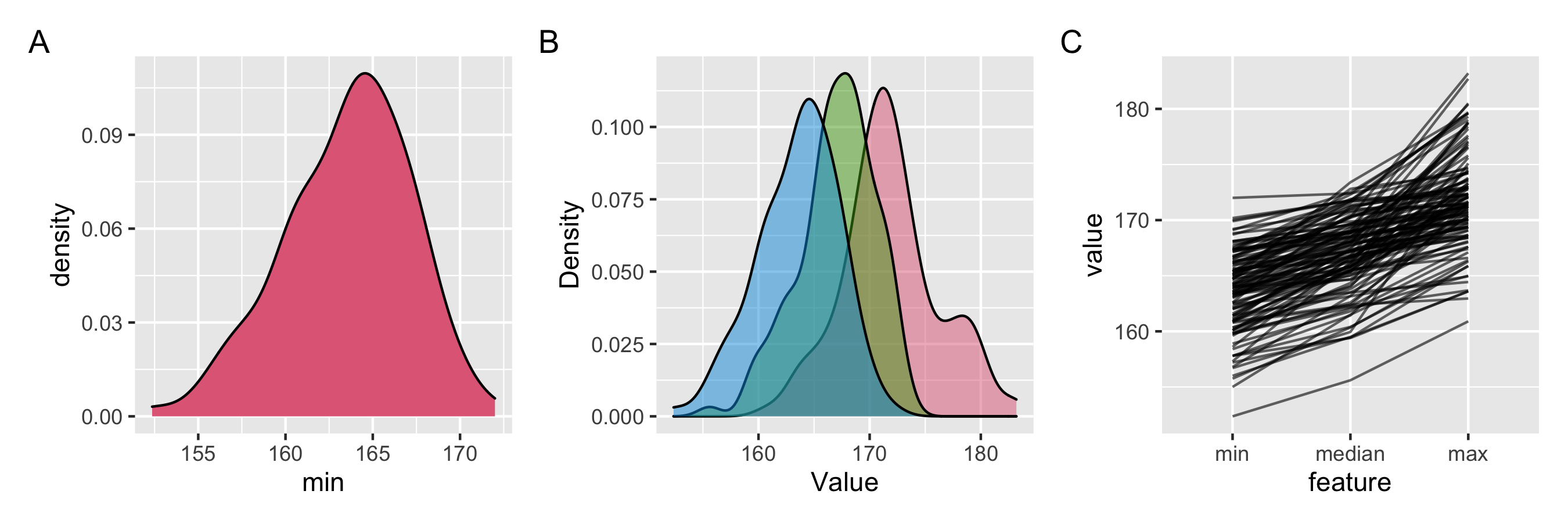} 

}

\caption[Three plots showing the distribution of minimum, median, and maximum values of height in centimeters]{Three plots showing the distribution of minimum, median, and maximum values of height in centimeters. Part A shows just the distribution of minimum, part B shows the distribution of minimum, median, and maximum, and part C shows these three values plotted together as a line graph. We see that there is overlap amongst all three statistics. That is, some countries minimum heights are taller that some countries maximum heights.}\label{fig:feature-min-med-max}
\end{figure}
\end{Schunk}

These sets of features can be pre-specified, for example,
\texttt{brolgar} provides a five number summary (minimum, 25th quantile,
median, mean, 75th quantile, and maximum) of the data with
\texttt{feat\_five\_num}:

\begin{Schunk}
\begin{Sinput}
heights_five <- heights_brolgar %>%
  features(height_cm, feat_five_num)

heights_five
\end{Sinput}
\begin{Soutput}
#> # A tibble: 119 x 6
#>    country       min   q25   med   q75   max
#>    <chr>       <dbl> <dbl> <dbl> <dbl> <dbl>
#>  1 Afghanistan  161.  164.  167.  168.  168.
#>  2 Algeria      166.  168.  169   170.  171.
#>  3 Angola       159.  160.  167.  168.  169.
#>  4 Argentina    167.  168.  168.  170.  174.
#>  5 Armenia      164.  166.  169.  172.  172.
#>  6 Australia    170   171.  172.  173.  178.
#>  7 Austria      162.  164.  167.  169.  179.
#>  8 Azerbaijan   170.  171.  172.  172.  172.
#>  9 Bangladesh   160.  162.  162.  163.  164.
#> 10 Belgium      163.  164.  166.  168.  177.
#> # ... with 109 more rows
\end{Soutput}
\end{Schunk}

This takes the \texttt{heights} data, pipes it to \texttt{features}, and
then instructs it to summarise the \texttt{height\_cm} variable, using
\texttt{feat\_five\_num}. There are several handy functions for
calculating features of the data that \texttt{brolgar} provides. These
all start with \texttt{feat\_}, and include:

\begin{itemize}
\tightlist
\item
  \texttt{feat\_ranges()}: min, max, range difference, interquartile
  range;
\item
  \texttt{feat\_spread()}: variance, standard deviation, median absolute
  distance, and interquartile range;
\item
  \texttt{feat\_monotonic()}: is it always increasing, decreasing, or
  unvarying?;
\item
  \texttt{feat\_diff\_summary()}: the summary statistics of the
  differences amongst a value, including the five number summary, as
  well as the standard deviation and variance;
\item
  \texttt{feat\_brolgar()}, which will calculate all features available
  in the \texttt{brolgar} package.
\item
  Other examples of features from the \texttt{feasts} package.
\end{itemize}

\hypertarget{feature-sets}{%
\subsubsection{Feature Sets}\label{feature-sets}}

If you want to run many or all features from a package on your data you
can collect them all with \texttt{feature\_set}. For example:

\begin{Schunk}
\begin{Sinput}
library(fabletools)
feat_set_brolgar <- feature_set(pkgs = "brolgar")
length(feat_set_brolgar)
\end{Sinput}
\begin{Soutput}
#> [1] 6
\end{Soutput}
\end{Schunk}

You could then run these like so:

\begin{Schunk}
\begin{Sinput}
heights_brolgar %>%
  features(height_cm, feat_set_brolgar)
\end{Sinput}
\begin{Soutput}
#> # A tibble: 119 x 46
#>    country min...1 med...2 max...3 min...4 q25...5 med...6 q75...7 max...8
#>    <chr>     <dbl>   <dbl>   <dbl>   <dbl>   <dbl>   <dbl>   <dbl>   <dbl>
#>  1 Afghan~    161.    167.    168.    161.    164.    167.    168.    168.
#>  2 Algeria    166.    169     171.    166.    168.    169     170.    171.
#>  3 Angola     159.    167.    169.    159.    160.    167.    168.    169.
#>  4 Argent~    167.    168.    174.    167.    168.    168.    170.    174.
#>  5 Armenia    164.    169.    172.    164.    166.    169.    172.    172.
#>  6 Austra~    170     172.    178.    170     171.    172.    173.    178.
#>  7 Austria    162.    167.    179.    162.    164.    167.    169.    179.
#>  8 Azerba~    170.    172.    172.    170.    171.    172.    172.    172.
#>  9 Bangla~    160.    162.    164.    160.    162.    162.    163.    164.
#> 10 Belgium    163.    166.    177.    163.    164.    166.    168.    177.
#> # ... with 109 more rows, and 37 more variables: min...9 <dbl>, max...10 <dbl>,
#> #   range_diff...11 <dbl>, iqr...12 <dbl>, var...13 <dbl>, sd...14 <dbl>,
#> #   mad...15 <dbl>, iqr...16 <dbl>, min...17 <dbl>, max...18 <dbl>,
#> #   median <dbl>, mean <dbl>, q25...21 <dbl>, q75...22 <dbl>, range1 <dbl>,
#> #   range2 <dbl>, range_diff...25 <dbl>, sd...26 <dbl>, var...27 <dbl>,
#> #   mad...28 <dbl>, iqr...29 <dbl>, increase...30 <dbl>, decrease...31 <dbl>,
#> #   unvary...32 <dbl>, diff_min <dbl>, diff_q25 <dbl>, diff_median <dbl>,
#> #   diff_mean <dbl>, diff_q75 <dbl>, diff_max <dbl>, diff_var <dbl>,
#> #   diff_sd <dbl>, diff_iqr <dbl>, increase...42 <lgl>, decrease...43 <lgl>,
#> #   unvary...44 <lgl>, monotonic <lgl>
\end{Soutput}
\end{Schunk}

To see other features available in the \texttt{feasts} R package run
\texttt{library(feasts)} then \texttt{?fabletools::feature\_set}.

\hypertarget{creating-your-own-feature}{%
\subsubsection{Creating Your Own
Feature}\label{creating-your-own-feature}}

To create your own features or summaries to pass to \texttt{features},
you provide a named vector of functions. These can include functions
that you have written yourself. For example, returning the first three
elements of a series, by writing our own \texttt{second} and
\texttt{third} functions.

\begin{Schunk}
\begin{Sinput}
second <- function(x) nth(x, n = 2)
third <- function(x) nth(x, n = 3)

feat_first_three <- c(first = first,
                      second = second,
                      third = third)
\end{Sinput}
\end{Schunk}

These are then passed to \texttt{features} like so:

\begin{Schunk}
\begin{Sinput}
heights_brolgar %>%
  features(height_cm, feat_first_three)
\end{Sinput}
\begin{Soutput}
#> # A tibble: 119 x 4
#>    country     first second third
#>    <chr>       <dbl>  <dbl> <dbl>
#>  1 Afghanistan  168.   166.  167.
#>  2 Algeria      169.   166.  169 
#>  3 Angola       160.   159.  160.
#>  4 Argentina    170.   168.  168 
#>  5 Armenia      169.   168.  166.
#>  6 Australia    170    171.  170.
#>  7 Austria      165.   163.  162.
#>  8 Azerbaijan   170.   171.  171.
#>  9 Bangladesh   162.   162.  164.
#> 10 Belgium      163.   164.  164 
#> # ... with 109 more rows
\end{Soutput}
\end{Schunk}

As well, \texttt{brolgar} provides some useful additional features for
the five number summary, \texttt{feat\_five\_num}, whether keys are
monotonically increasing \texttt{feat\_monotonic}, and measures of
spread or variation, \texttt{feat\_spread}. Inside \texttt{brolgar}, the
features are created with the following syntax:

\begin{Schunk}
\begin{Sinput}
feat_five_num <- function(x, ...) {
  c(
    min = b_min(x, ...),
    q25 = b_q25(x, ...),
    med = b_median(x, ...),
    q75 = b_q75(x, ...),
    max = b_max(x, ...)
  )
}
\end{Sinput}
\end{Schunk}

Here the functions \texttt{b\_} are functions with a default of
\texttt{na.rm\ =\ TRUE}, and in the cases of quantiles, they use
\texttt{type\ =\ 8}, and \texttt{names\ =\ FALSE}. What is particularly
useful is that these will work on any type of time series data, and you
can use other more typical time series features from the \texttt{feasts}
package, such as autocorrelation, \texttt{feat\_acf()} and Seasonal and
Trend decomposition using Loess \texttt{feat\_stl()} \citep{feasts}.

This demonstrates a workflow that can be used to understand and explore
your longitudinal data. The \texttt{brolgar} package builds upon this
workflow made available by \texttt{feasts} and \texttt{fabletools}.
Users can also create their own features to summarise the data.

\hypertarget{breaking-up-the-spaghetti}{%
\section{Breaking up the Spaghetti}\label{breaking-up-the-spaghetti}}

Plots like Figure \ref{fig:spaghetti} are often called, ``spaghetti
plots'', and can be useful for a high level understanding as a whole.
However, we cannot process and understand the individuals when the data
is presented like this.

\hypertarget{sampling}{%
\subsection{Sampling}\label{sampling}}

Just how spaghetti is portioned out for consumption, we can sample some
of the data by randomly sampling the data into sub-plots with the
\texttt{facet\_sample()} function (Figure \ref{fig:facet-sample}).

\begin{Schunk}
\begin{Sinput}
ggplot(heights_brolgar,
       aes(x = year,
           y = height_cm,
           group = country)) + 
  geom_line() + 
  facet_sample() + 
  scale_x_continuous(breaks = c(1750, 1850, 1950))
\end{Sinput}
\begin{figure}

{\centering \includegraphics[width=0.95\linewidth]{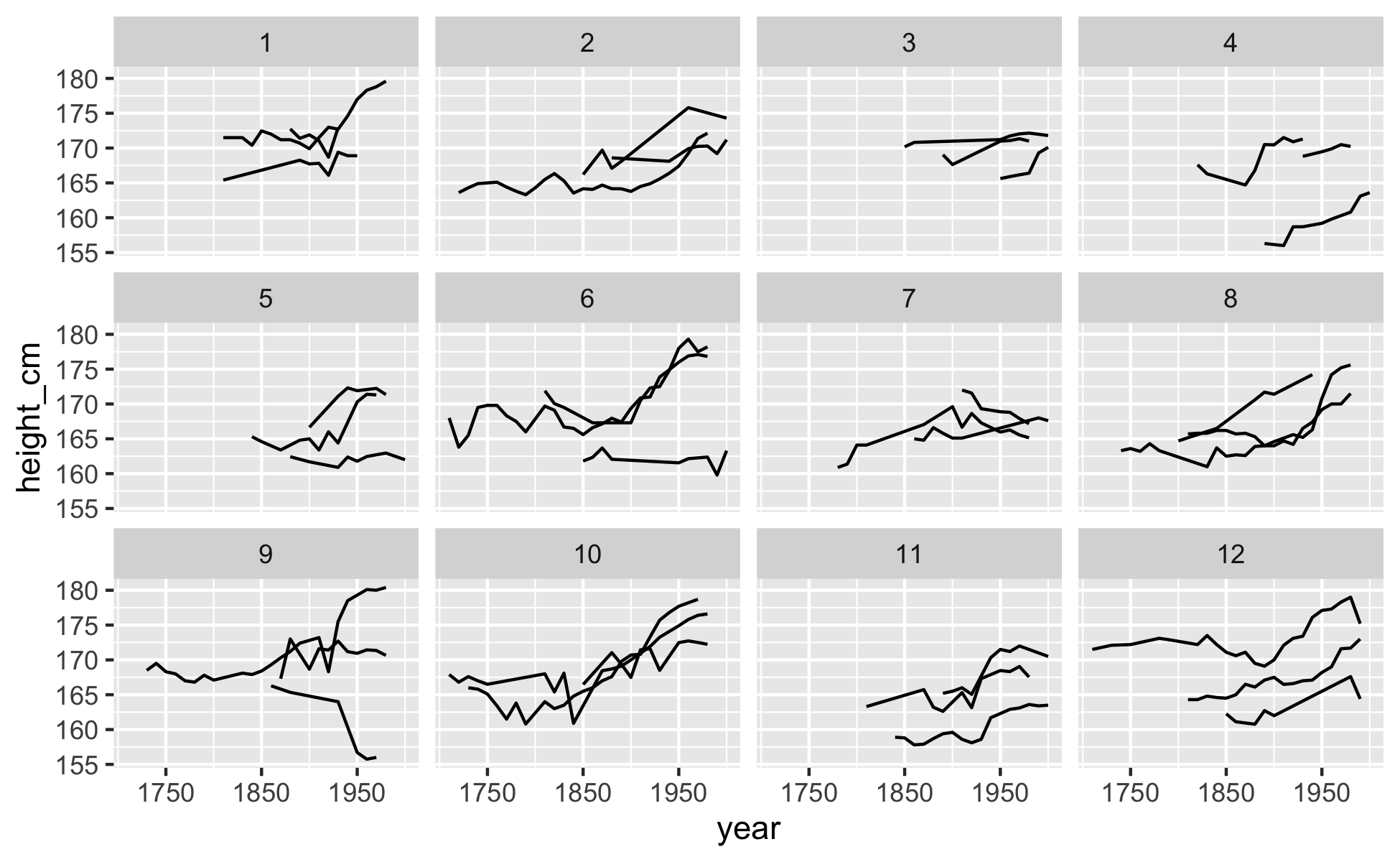} 

}

\caption[Twelve facets with three keys per facet shown]{Twelve facets with three keys per facet shown. This allows us to quickly view a random sample of the data.}\label{fig:facet-sample}
\end{figure}
\end{Schunk}

This defaults to 12 facets and 3 samples per facet, and provides options
for the number of facets, and the number of samples per facet. This
means the user only needs to consider the most relevant questions: ``How
many keys per facet?'' and ``How many facets to look at?''. The code to
change the figure from Figure \ref{fig:spaghetti} into
\ref{fig:facet-sample} requires only one line of code, shown below:

\begin{verbatim}
ggplot(heights_brolgar,
       aes(x = year,
           y = height_cm,
           group = country)) + 
  geom_line() + 
  facet_sample()
\end{verbatim}

\hypertarget{stratifying}{%
\subsection{Stratifying}\label{stratifying}}

Extending this idea of samples, we can instead look at \textbf{all} of
the data, spread out equally over facets, using
\texttt{facet\_strata()}. It uses 12 facets by default, controllable
with \texttt{n\_strata}. The code to do so is shown below, creating
Figure \ref{fig:facet-strata}.

\begin{Schunk}
\begin{Sinput}
ggplot(heights_brolgar,
       aes(x = year,
           y = height_cm,
           group = country)) + 
  geom_line() + 
  facet_strata() +
  scale_x_continuous(breaks = c(1750, 1850, 1950))
\end{Sinput}
\begin{figure}

{\centering \includegraphics[width=0.95\linewidth]{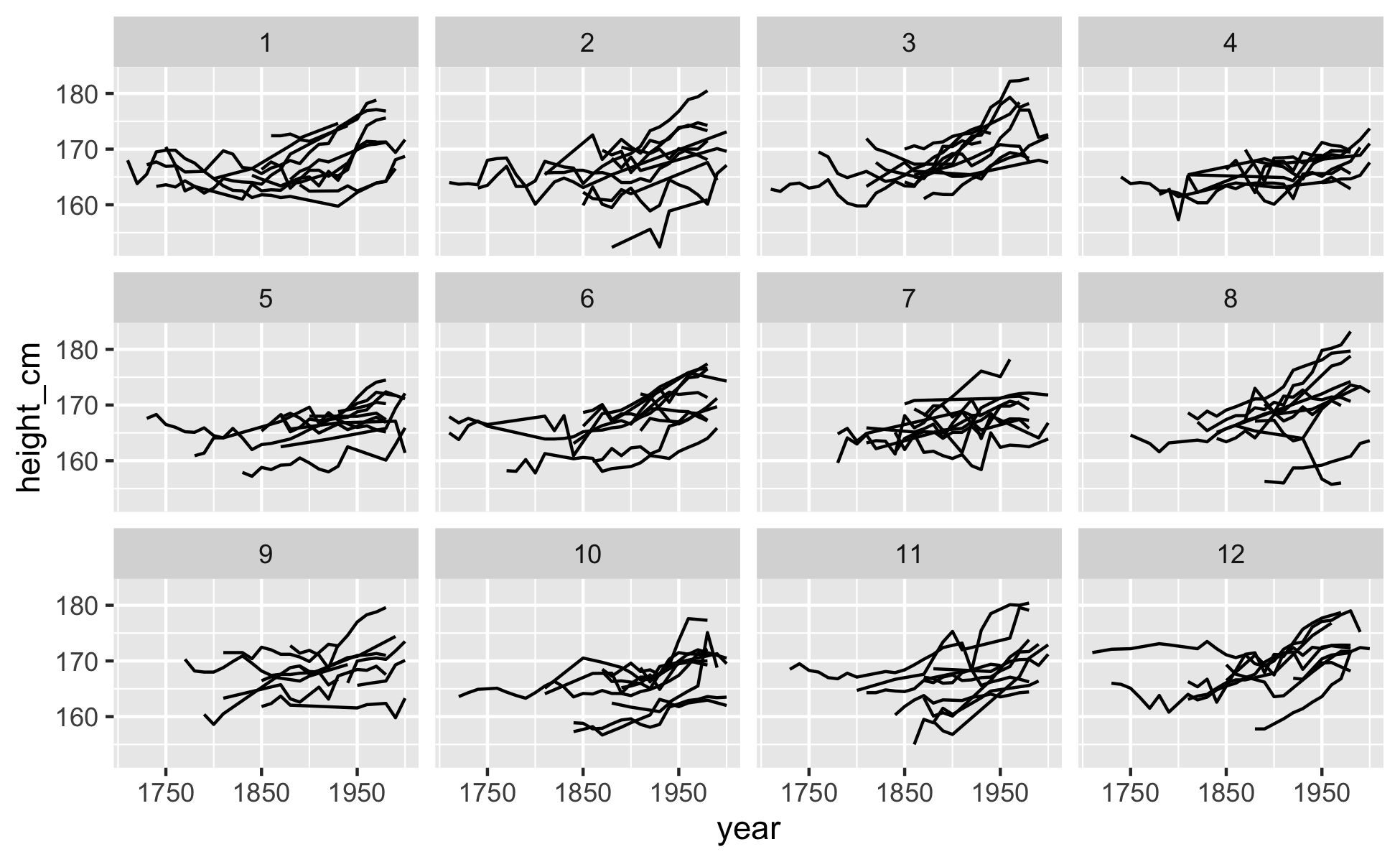} 

}

\caption[All of the data is shown by spreading out each key across twelve facets]{All of the data is shown by spreading out each key across twelve facets. Each key is only shown once, and is randomly allocated to a facet.}\label{fig:facet-strata}
\end{figure}
\end{Schunk}

\hypertarget{featuring}{%
\subsection{Featuring}\label{featuring}}

Figure \ref{fig:facet-strata} and Figure \ref{fig:facet-sample} only
show each key once, being randomly assigned to a facet. We can
meaningfully place the keys into facets, by arranging the heights
``along'' a variable, like \texttt{year}, using the \texttt{along}
argument in \texttt{facet\_strata} to produce Figure
\ref{fig:facet-year-along}:

\begin{Schunk}
\begin{Sinput}
ggplot(heights_brolgar,
       aes(x = year,
           y = height_cm,
           group = country)) + 
  geom_line() + 
  facet_strata(along = -year) + 
  scale_x_continuous(breaks = c(1750, 1850, 1950))
\end{Sinput}
\begin{figure}

{\centering \includegraphics[width=0.95\linewidth]{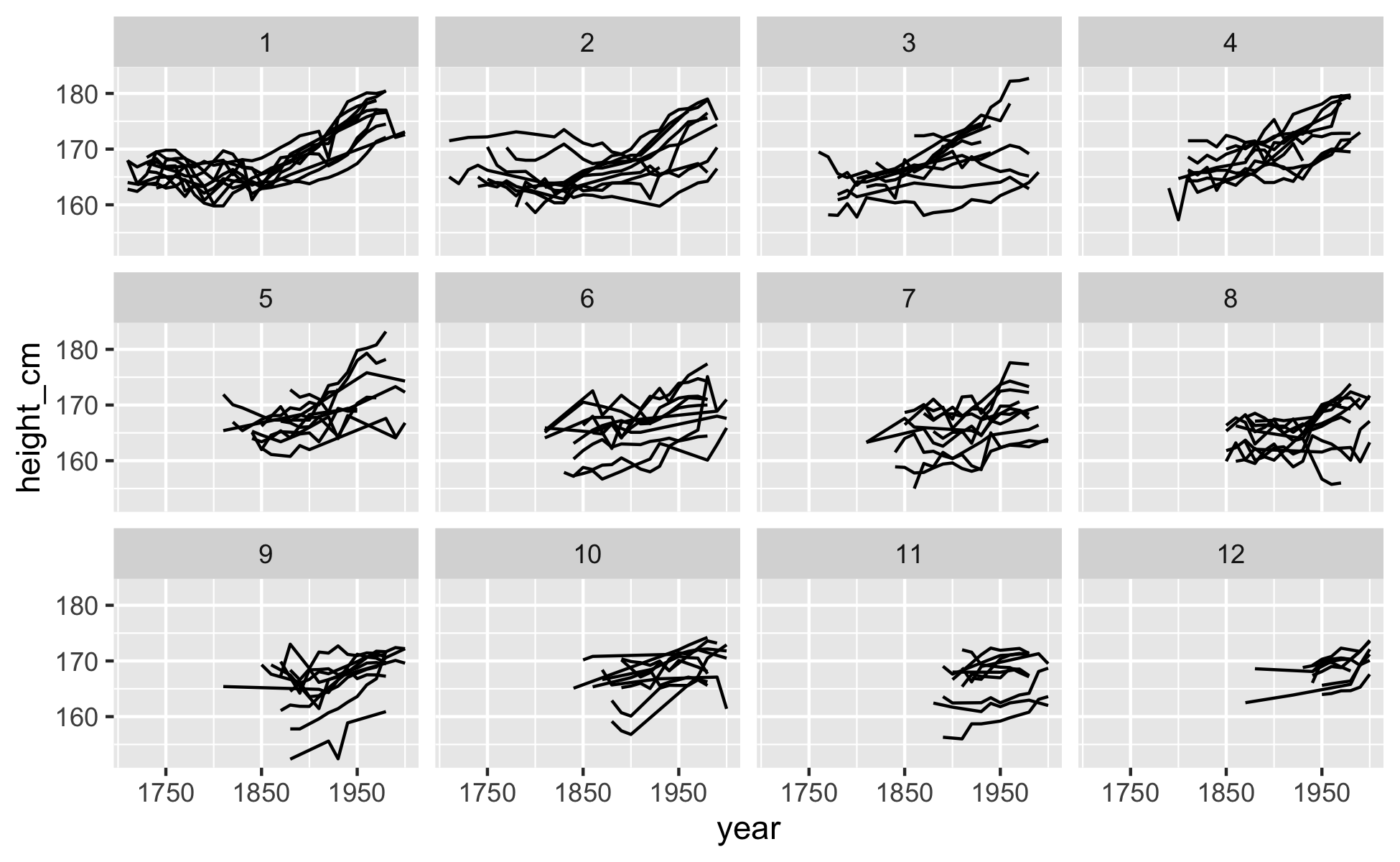} 

}

\caption[Displaying all the data across twelve facets]{Displaying all the data across twelve facets. Instead of each key being randomly in a facet, each facet displays a specified range of values of year. In this case, the top left facet shows the keys with the earliest starting year, and the bottom right shows the facet with the latest starting year.}\label{fig:facet-year-along}
\end{figure}
\end{Schunk}

We have not lost any of the data, only the order in which they are
presented has changed. We learn the distribution and changes in heights
over time, and those measured from the earliest times appear to be more
similar, but there is much wider variation in the middle years, and then
for more recent heights measured from the early 1900s, the heights are
more similar. The starting point of each of these years seems to
increase at roughly the same interval. This informs us that the starting
times of the years is approximately uniform.

Together \texttt{facet\_sample()} and \texttt{facet\_strata()} allow for
rapid exploration, by focusing on relevant questions instead of the
minutiae. This is achieved by appropriately randomly assigning while
maintaining key structure, keeping the correct number of keys per plot,
and so on. For example, \texttt{facet\_sample()} the questions are:
``How many lines per facet'' and ``How many facets?'', and for
\texttt{facet\_strata()} the questions are: ``How many facets /
strata?'' and ``What to arrange plots along?''.

Answering these questions keeps the analysis in line with the analytic
goals of exploring the data, rather than distracting to minutiae. This
is a key theme of improving tools for data analysis. Abstracting away
the parts that are not needed, so the analyst can focus on the task at
hand.

Under the hood, \texttt{facet\_sample()} and \texttt{facet\_strata()}
are powered with \texttt{sample\_n\_keys()} and
\texttt{stratify\_keys()}. These can be used to create data structures
used in \texttt{facet\_sample()} and \texttt{facet\_strata()}, and
extend them for other purposes.

Using a \texttt{tsibble} stores important key and index components, in
turn allowing for better ways to break up spaghetti plots so we can look
at many and all sub-samples using \texttt{facet\_sample()} and
\texttt{facet\_strata()}.

\hypertarget{book-keeping}{%
\section{Book-keeping}\label{book-keeping}}

Longitudinal data is not always measured at the same time and at the
same frequency. When exploring longitudinal data, a useful first step is
to explore the frequency of measurements of the index. We can check if
the index is regular using \texttt{index\_regular()} and summarise the
spacing of the index with \texttt{index\_summary()}. These are S3
methods, so for \texttt{data.frame} objects, the \texttt{index} must be
specified, however for the \texttt{tsibble} objects, the defined index
is used.

\begin{Schunk}
\begin{Sinput}
index_summary(heights_brolgar)
\end{Sinput}
\begin{Soutput}
#>    Min. 1st Qu.  Median    Mean 3rd Qu.    Max. 
#>    1710    1782    1855    1855    1928    2000
\end{Soutput}
\begin{Sinput}
index_regular(heights_brolgar)
\end{Sinput}
\begin{Soutput}
#> [1] TRUE
\end{Soutput}
\end{Schunk}

We can explore how many observations per country by counting the number
of observations with \texttt{features}, like so:

\begin{Schunk}
\begin{Sinput}
heights_brolgar %>% features(year, n_obs)
\end{Sinput}
\begin{Soutput}
#> # A tibble: 119 x 2
#>    country     n_obs
#>    <chr>       <int>
#>  1 Afghanistan     5
#>  2 Algeria         5
#>  3 Angola          9
#>  4 Argentina      20
#>  5 Armenia        11
#>  6 Australia      10
#>  7 Austria        18
#>  8 Azerbaijan      7
#>  9 Bangladesh      9
#> 10 Belgium        10
#> # ... with 109 more rows
\end{Soutput}
\end{Schunk}

This can be further summarised by counting the number of times there are
a given number of observations:

\begin{Schunk}
\begin{Sinput}
heights_brolgar %>% features(year, n_obs) %>% count(n_obs)
\end{Sinput}
\begin{Soutput}
#> # A tibble: 24 x 2
#>    n_obs     n
#>    <int> <int>
#>  1     5    11
#>  2     6    11
#>  3     7    13
#>  4     8     5
#>  5     9    12
#>  6    10    12
#>  7    11     9
#>  8    12     4
#>  9    13     7
#> 10    14     6
#> # ... with 14 more rows
\end{Soutput}
\end{Schunk}

Because we are exploring the temporal patterns, we cannot reliably say
anything about those individuals with few measurements. The data used,
\texttt{heights\_brolgar} has less than 5 measurements. This was done
using \texttt{add\_n\_obs()}, which adds the number of observations to
the existing data. Overall this drops 25 countries, leaves us with 119
out of the original 144 countries.

\begin{Schunk}
\begin{Sinput}
heights_brolgar <- heights %>% 
  add_n_obs() %>% 
  filter(n_obs >= 5)
\end{Sinput}
\end{Schunk}

We can further explore when countries are first being measured using
\texttt{features} to find the first year for each country number of
starting years with the \texttt{first} function from \texttt{dplyr}, and
explore this with a visualisation (Figure \ref{fig:feature-first-gg}).

\begin{Schunk}
\begin{Sinput}
heights_brolgar %>% 
  features(year, c(first = first))
\end{Sinput}
\begin{Soutput}
#> # A tibble: 119 x 2
#>    country     first
#>    <chr>       <dbl>
#>  1 Afghanistan  1870
#>  2 Algeria      1910
#>  3 Angola       1790
#>  4 Argentina    1770
#>  5 Armenia      1850
#>  6 Australia    1850
#>  7 Austria      1750
#>  8 Azerbaijan   1850
#>  9 Bangladesh   1850
#> 10 Belgium      1810
#> # ... with 109 more rows
\end{Soutput}
\end{Schunk}

\begin{Schunk}
\begin{Sinput}
heights_brolgar %>% 
  features(year, c(first = first)) %>% 
  ggplot(aes(x = first)) +
  geom_bar()
\end{Sinput}
\begin{figure}

{\centering \includegraphics[width=0.6\linewidth]{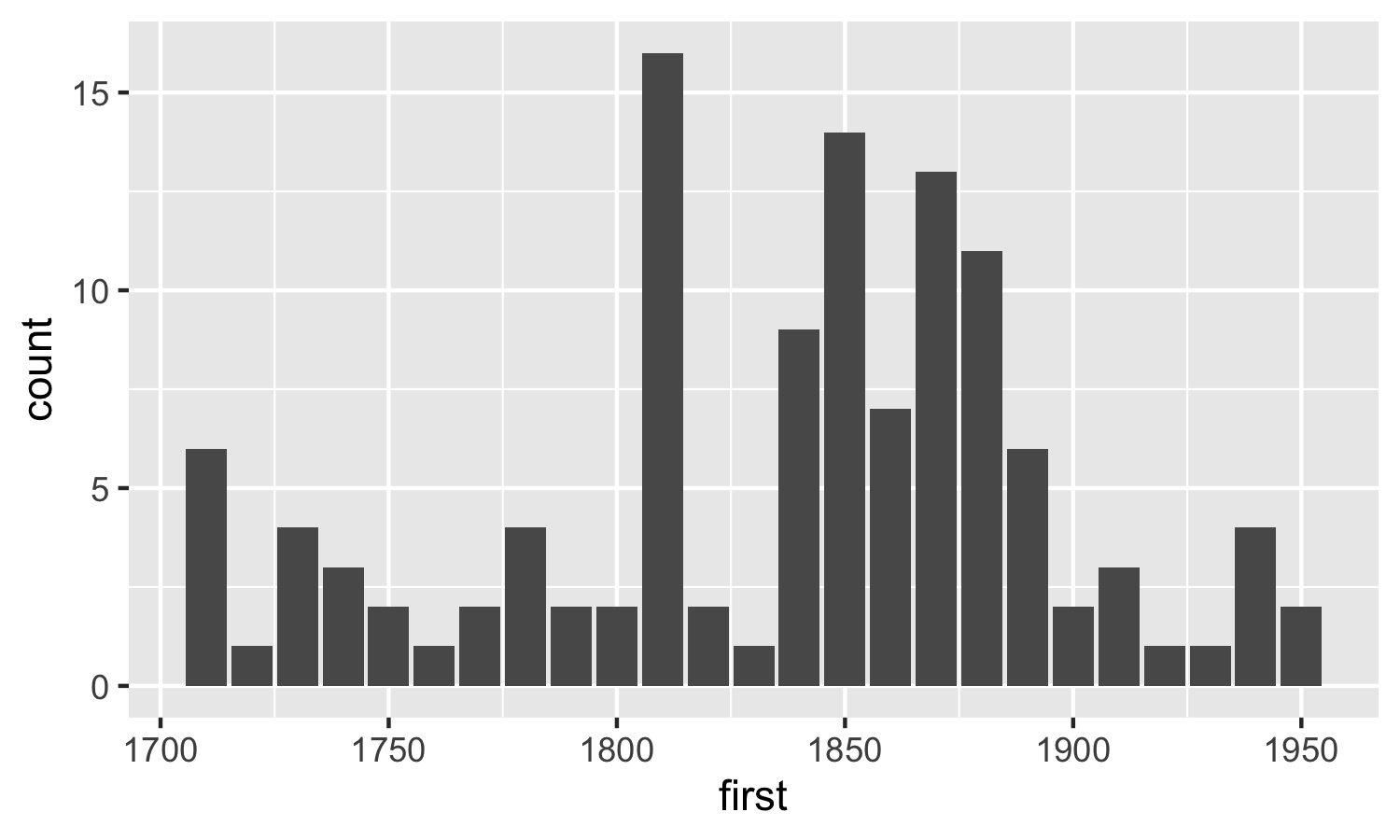} 

}

\caption[Distribution of starting years of measurement]{Distribution of starting years of measurement. The data is already binned into 10 year blocks. Most of the years start between 1840 and 1900.}\label{fig:feature-first-gg}
\end{figure}
\end{Schunk}

We can explore the variation in first year using
\texttt{feat\_diff\_summary}. This combines many summaries of the
differences in \texttt{year}.

\begin{Schunk}
\begin{Sinput}
heights_diffs <- heights_brolgar %>% 
  features(year, feat_diff_summary)

heights_diffs
\end{Sinput}
\begin{Soutput}
#> # A tibble: 119 x 10
#>    country diff_min diff_q25 diff_median diff_mean diff_q75 diff_max diff_var
#>    <chr>      <dbl>    <dbl>       <dbl>     <dbl>    <dbl>    <dbl>    <dbl>
#>  1 Afghan~       10       10          30      32.5     55.8       60    692. 
#>  2 Algeria       10       10          10      22.5     39.2       60    625  
#>  3 Angola        10       10          10      17.5     10         70    450  
#>  4 Argent~       10       10          10      11.6     10         40     47.4
#>  5 Armenia       10       10          10      15       20.8       30     72.2
#>  6 Austra~       10       10          10      13.3     10         40    100  
#>  7 Austria       10       10          10      13.5     10         40     74.3
#>  8 Azerba~       10       10          10      25       25.8       90   1030  
#>  9 Bangla~       10       10          10      18.8     15.8       70    441. 
#> 10 Belgium       10       10          10      16.7     23.3       40    125  
#> # ... with 109 more rows, and 2 more variables: diff_sd <dbl>, diff_iqr <dbl>
\end{Soutput}
\end{Schunk}

This is particularly useful as using \texttt{diff} on \texttt{year}
would return a very wide dataset that is hard to explore:

\begin{Schunk}
\begin{Sinput}
heights_brolgar %>% 
  features(year, diff)
\end{Sinput}
\begin{Soutput}
#> # A tibble: 119 x 30
#>    country  ...1  ...2  ...3  ...4  ...5  ...6  ...7  ...8  ...9 ...10 ...11
#>    <chr>   <dbl> <dbl> <dbl> <dbl> <dbl> <dbl> <dbl> <dbl> <dbl> <dbl> <dbl>
#>  1 Afghan~    10    50    60    10    NA    NA    NA    NA    NA    NA    NA
#>  2 Algeria    10    10    60    10    NA    NA    NA    NA    NA    NA    NA
#>  3 Angola     10    10    70    10    10    10    10    10    NA    NA    NA
#>  4 Argent~    10    10    10    10    10    10    10    10    10    10    10
#>  5 Armenia    10    30    10    10    30    20    10    10    10    10    NA
#>  6 Austra~    10    10    10    10    10    10    10    40    10    NA    NA
#>  7 Austria    20    10    10    30    10    10    10    10    10    10    10
#>  8 Azerba~    10    90    10    10    10    20    NA    NA    NA    NA    NA
#>  9 Bangla~    10    10    10    70    10    20    10    10    NA    NA    NA
#> 10 Belgium    10    10    10    10    10    10    30    40    20    NA    NA
#> # ... with 109 more rows, and 18 more variables: ...12 <dbl>, ...13 <dbl>,
#> #   ...14 <dbl>, ...15 <dbl>, ...16 <dbl>, ...17 <dbl>, ...18 <dbl>,
#> #   ...19 <dbl>, ...20 <dbl>, ...21 <dbl>, ...22 <dbl>, ...23 <dbl>,
#> #   ...24 <dbl>, ...25 <dbl>, ...26 <dbl>, ...27 <dbl>, ...28 <dbl>,
#> #   ...29 <dbl>
\end{Soutput}
\end{Schunk}

We can then look at the summaries of the differences in year by changing
to long form and facetting (Figure \ref{fig:heights-long-feat-diff}), we
learn about the range of intervals between measurements, the smallest
being 10 years, the largest being 125, and that most of the data is
measured between 10 and 30 years.

\begin{Schunk}
\begin{figure}

{\centering \includegraphics[width=0.95\linewidth]{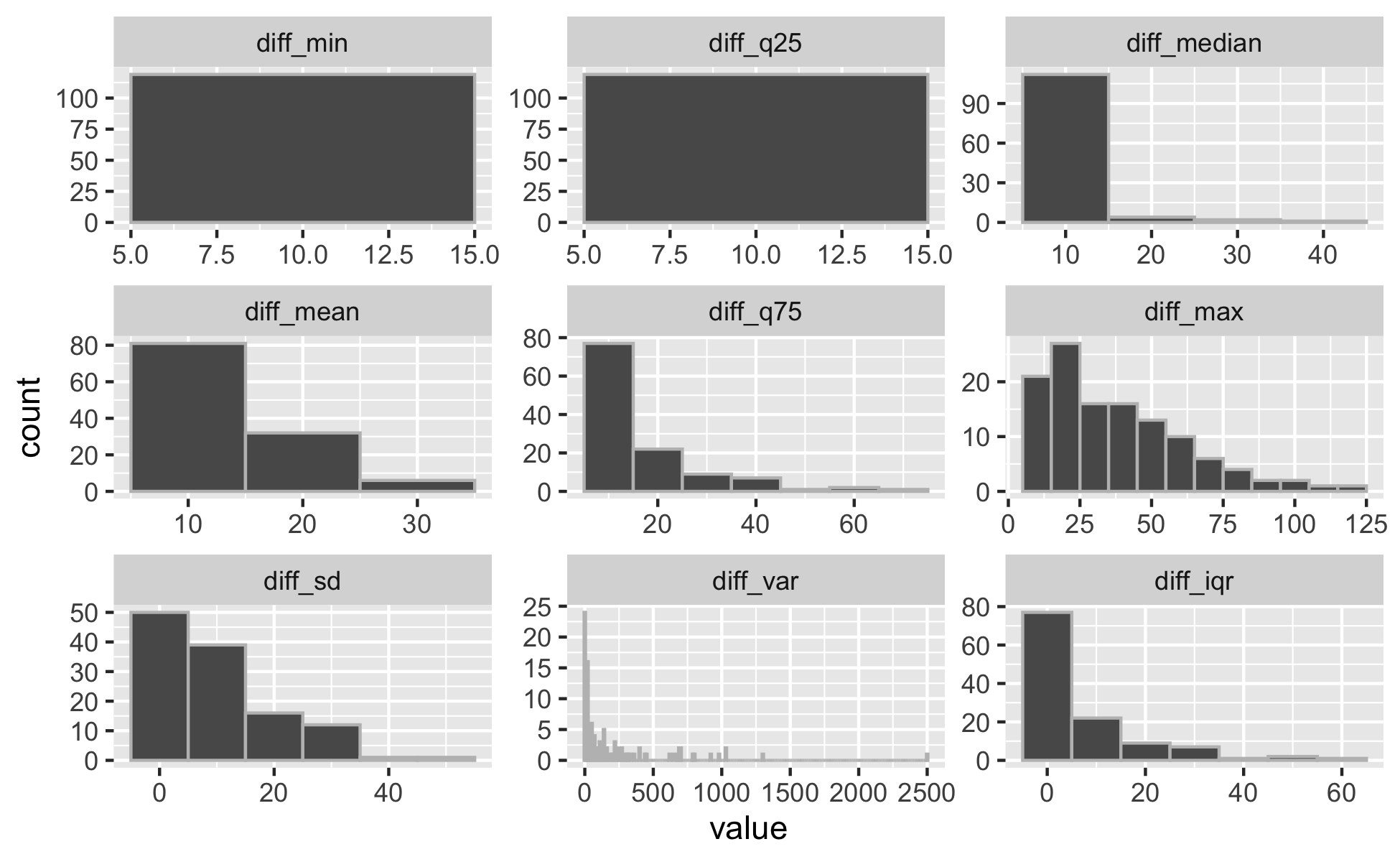} 

}

\caption[Exploring the different summary statistics of the differences amongst the years]{Exploring the different summary statistics of the differences amongst the years. We learn that the smallest interval between measurements is 10 years, and the largest interval is between 10 and 125 years, and that most of the data is measured between 10 and 30 or so years.}\label{fig:heights-long-feat-diff}
\end{figure}
\end{Schunk}

\hypertarget{finding-waldo}{%
\section{Finding Waldo}\label{finding-waldo}}

Looking at a spaghetti plot, it can be hard to identify which lines are
the most interesting, or unusual. A workflow to identify interesting
individuals to start with is given below:

\begin{enumerate}
\def\labelenumi{\arabic{enumi}.}
\tightlist
\item
  Decide upon an interesting feature (e.g., maximum)
\item
  This feature produces one value per key
\item
  Examine the distribution of the feature
\item
  Join this table back to the data to get all observations for those
  keys
\item
  Arrange the keys or filter, using the feature
\item
  Display the data for selected keys
\end{enumerate}

This workflow is now demonstrated. Firstly, we \textbf{deicde on an
interesting feature}, ``maximum height'', and whether height is always
increasing. We calculate our own ``feature'', calculating maximum
height, and whether a value is increasing (with brolgar's
\texttt{increasing} function) as follows:

\begin{Schunk}
\begin{Sinput}
heights_max_in <- heights_brolgar %>% 
  features(height_cm, list(max = max,
                           increase = increasing))

heights_max_in
\end{Sinput}
\begin{Soutput}
#> # A tibble: 119 x 3
#>    country       max increase
#>    <chr>       <dbl> <lgl>   
#>  1 Afghanistan  168. FALSE   
#>  2 Algeria      171. FALSE   
#>  3 Angola       169. FALSE   
#>  4 Argentina    174. FALSE   
#>  5 Armenia      172. FALSE   
#>  6 Australia    178. FALSE   
#>  7 Austria      179. FALSE   
#>  8 Azerbaijan   172. FALSE   
#>  9 Bangladesh   164. FALSE   
#> 10 Belgium      177. FALSE   
#> # ... with 109 more rows
\end{Soutput}
\end{Schunk}

This returns a dataset of \textbf{one value per key}. Figure
\ref{fig:heights-max-in-examine} \textbf{examines the distribution of
the features}, showing us the distribution of maximum height, and the
number of countries that are always increasing.

\begin{Schunk}
\begin{figure}

{\centering \includegraphics[width=0.95\linewidth]{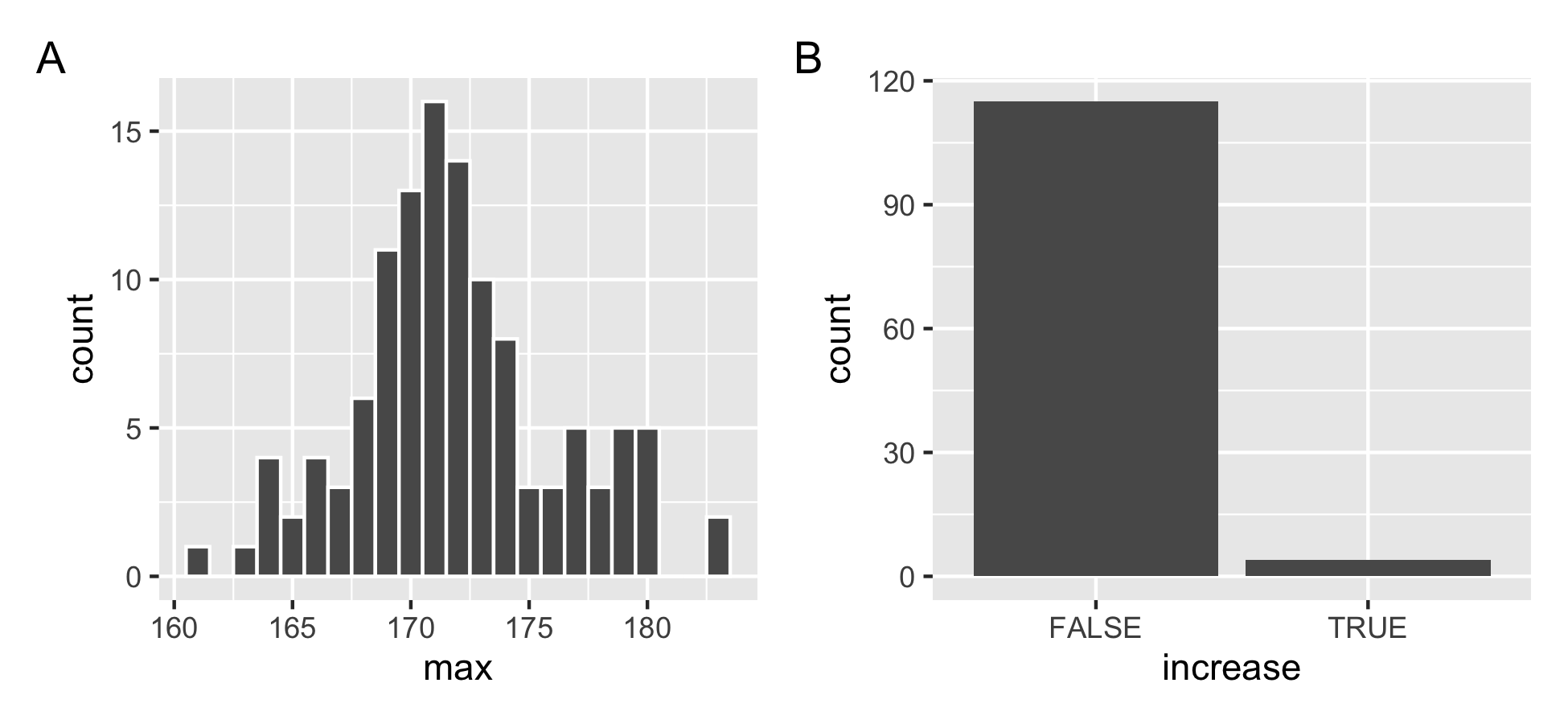} 

}

\caption[The different distributions of the features - A depicting the distribution of maximum height, and B displaying the number of countries that are always increasing (FALSE), and always increasing (TRUE)]{The different distributions of the features - A depicting the distribution of maximum height, and B displaying the number of countries that are always increasing (FALSE), and always increasing (TRUE). We note that the average maximum heights range from about 160cm to 185cm, with most being around 170cm. We also learn that the vast majority of countries are not always increasing in height through time.}\label{fig:heights-max-in-examine}
\end{figure}
\end{Schunk}

We can now \textbf{join this table back to the data to get all
observations for those keys} to move from one key per row to all many
rows per key.

\begin{Schunk}
\begin{Sinput}
heights_max_in_full <- heights_max_in %>% 
  left_join(heights_brolgar,
            by = "country")

heights_max_in_full
\end{Sinput}
\begin{Soutput}
#> # A tibble: 1,406 x 9
#>    country       max increase  year n_obs continent height_cm year0 country_fct
#>    <chr>       <dbl> <lgl>    <dbl> <int> <chr>         <dbl> <dbl> <fct>      
#>  1 Afghanistan  168. FALSE     1870     5 Asia           168.   160 Afghanistan
#>  2 Afghanistan  168. FALSE     1880     5 Asia           166.   170 Afghanistan
#>  3 Afghanistan  168. FALSE     1930     5 Asia           167.   220 Afghanistan
#>  4 Afghanistan  168. FALSE     1990     5 Asia           167.   280 Afghanistan
#>  5 Afghanistan  168. FALSE     2000     5 Asia           161.   290 Afghanistan
#>  6 Algeria      171. FALSE     1910     5 Africa         169.   200 Algeria    
#>  7 Algeria      171. FALSE     1920     5 Africa         166.   210 Algeria    
#>  8 Algeria      171. FALSE     1930     5 Africa         169    220 Algeria    
#>  9 Algeria      171. FALSE     1990     5 Africa         171.   280 Algeria    
#> 10 Algeria      171. FALSE     2000     5 Africa         170.   290 Algeria    
#> # ... with 1,396 more rows
\end{Soutput}
\end{Schunk}

We can then \textbf{arrange the keys or filter, using the feature}, for
example, filtering only those countries that are only increasing:

\begin{Schunk}
\begin{Sinput}
heights_increase <- heights_max_in_full %>% filter(increase)
heights_increase
\end{Sinput}
\begin{Soutput}
#> # A tibble: 22 x 9
#>    country    max increase  year n_obs continent height_cm year0 country_fct
#>    <chr>    <dbl> <lgl>    <dbl> <int> <chr>         <dbl> <dbl> <fct>      
#>  1 Honduras  168. TRUE      1950     6 Americas       164.   240 Honduras   
#>  2 Honduras  168. TRUE      1960     6 Americas       164.   250 Honduras   
#>  3 Honduras  168. TRUE      1970     6 Americas       165.   260 Honduras   
#>  4 Honduras  168. TRUE      1980     6 Americas       165.   270 Honduras   
#>  5 Honduras  168. TRUE      1990     6 Americas       165.   280 Honduras   
#>  6 Honduras  168. TRUE      2000     6 Americas       168.   290 Honduras   
#>  7 Moldova   174. TRUE      1840     5 Europe         165.   130 Moldova    
#>  8 Moldova   174. TRUE      1950     5 Europe         172.   240 Moldova    
#>  9 Moldova   174. TRUE      1960     5 Europe         173.   250 Moldova    
#> 10 Moldova   174. TRUE      1970     5 Europe         174.   260 Moldova    
#> # ... with 12 more rows
\end{Soutput}
\end{Schunk}

Or tallest country

\begin{Schunk}
\begin{Sinput}
heights_top <- heights_max_in_full %>% top_n(n = 1, wt = max)
heights_top
\end{Sinput}
\begin{Soutput}
#> # A tibble: 16 x 9
#>    country   max increase  year n_obs continent height_cm year0 country_fct
#>    <chr>   <dbl> <lgl>    <dbl> <int> <chr>         <dbl> <dbl> <fct>      
#>  1 Denmark  183. FALSE     1820    16 Europe         167.   110 Denmark    
#>  2 Denmark  183. FALSE     1830    16 Europe         165.   120 Denmark    
#>  3 Denmark  183. FALSE     1850    16 Europe         167.   140 Denmark    
#>  4 Denmark  183. FALSE     1860    16 Europe         168.   150 Denmark    
#>  5 Denmark  183. FALSE     1870    16 Europe         168.   160 Denmark    
#>  6 Denmark  183. FALSE     1880    16 Europe         170.   170 Denmark    
#>  7 Denmark  183. FALSE     1890    16 Europe         169.   180 Denmark    
#>  8 Denmark  183. FALSE     1900    16 Europe         170.   190 Denmark    
#>  9 Denmark  183. FALSE     1910    16 Europe         170    200 Denmark    
#> 10 Denmark  183. FALSE     1920    16 Europe         174.   210 Denmark    
#> 11 Denmark  183. FALSE     1930    16 Europe         174.   220 Denmark    
#> 12 Denmark  183. FALSE     1940    16 Europe         176.   230 Denmark    
#> 13 Denmark  183. FALSE     1950    16 Europe         180.   240 Denmark    
#> 14 Denmark  183. FALSE     1960    16 Europe         180.   250 Denmark    
#> 15 Denmark  183. FALSE     1970    16 Europe         181.   260 Denmark    
#> 16 Denmark  183. FALSE     1980    16 Europe         183.   270 Denmark
\end{Soutput}
\end{Schunk}

We can then display the data by highlighting it in the background, first
creating a background plot and overlaying the plots on top of this as an
additional ggplot layer, in Figure \ref{fig:gg-increase-max}.

\begin{Schunk}
\begin{figure}

{\centering \includegraphics[width=0.95\linewidth]{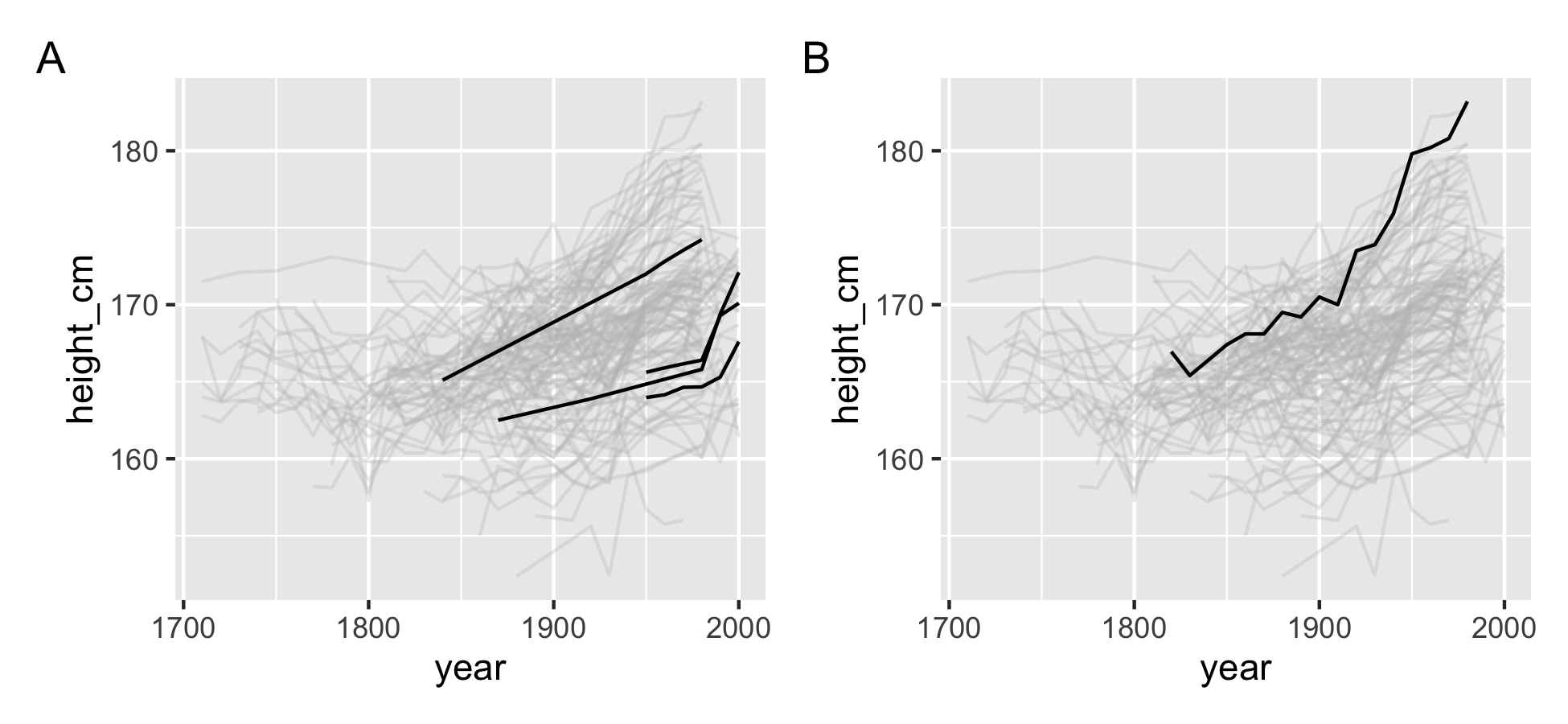} 

}

\caption[Plots of the data in the background, with the countries that always increase in height through time in A, and the country with the tallest people in B]{Plots of the data in the background, with the countries that always increase in height through time in A, and the country with the tallest people in B}\label{fig:gg-increase-max}
\end{figure}
\end{Schunk}

\hypertarget{dancing-with-models}{%
\section{Dancing with Models}\label{dancing-with-models}}

These same workflows can be used to interpret and explore a model. As
the data tends to follow a non linear trajectory, we use a general
additive model (gam) with the \texttt{mgcv} R package \citep{mgcv} using
the code below:

\begin{Schunk}
\begin{Sinput}
heights_gam <- gam(
    height_cm ~ s(year0, by = country_fct) + country_fct,
    data = heights_brolgar,
    method = "REML"
  )
\end{Sinput}
\end{Schunk}

This fits height in centimetres with a smooth effect for year for each
country, with a different intercept for each country. It is roughly
equivalent to a random intercept varying slope model. Note that this gam
model took approximately 8074 seconds to fit. We add the predicted and
residual values for the model below, as well as the residual sums of
squares for each country.

\begin{Schunk}
\begin{Sinput}
library(mgcv)
library(modelr)
heights_aug <- heights_brolgar %>%
  add_predictions(heights_gam, var = "pred") %>%
  add_residuals(heights_gam, var = "res") %>% 
  group_by_key() %>% 
  mutate(rss = sum(res^2)) %>% 
  ungroup()
\end{Sinput}
\end{Schunk}

We can use the previous approach to explore the model results. We can
take a look at a sample of the predictions along with the data, by using
\texttt{sample\_n\_keys}. This provides a useful way to explore some set
of the model predictions. In order to find those predictions that best
summarise the best, and worst, and in between, we need to use the
methods in the next section, ``stereotyping''.

\begin{Schunk}
\begin{Sinput}
heights_aug %>% 
  sample_n_keys(12) %>% 
  ggplot(aes(x = year,
             y = pred,
             group = country)) + 
  geom_line(colour = "steelblue") +
  geom_point(aes(y = height_cm)) + 
  facet_wrap(~country)
\end{Sinput}
\begin{figure}

{\centering \includegraphics[width=0.95\linewidth]{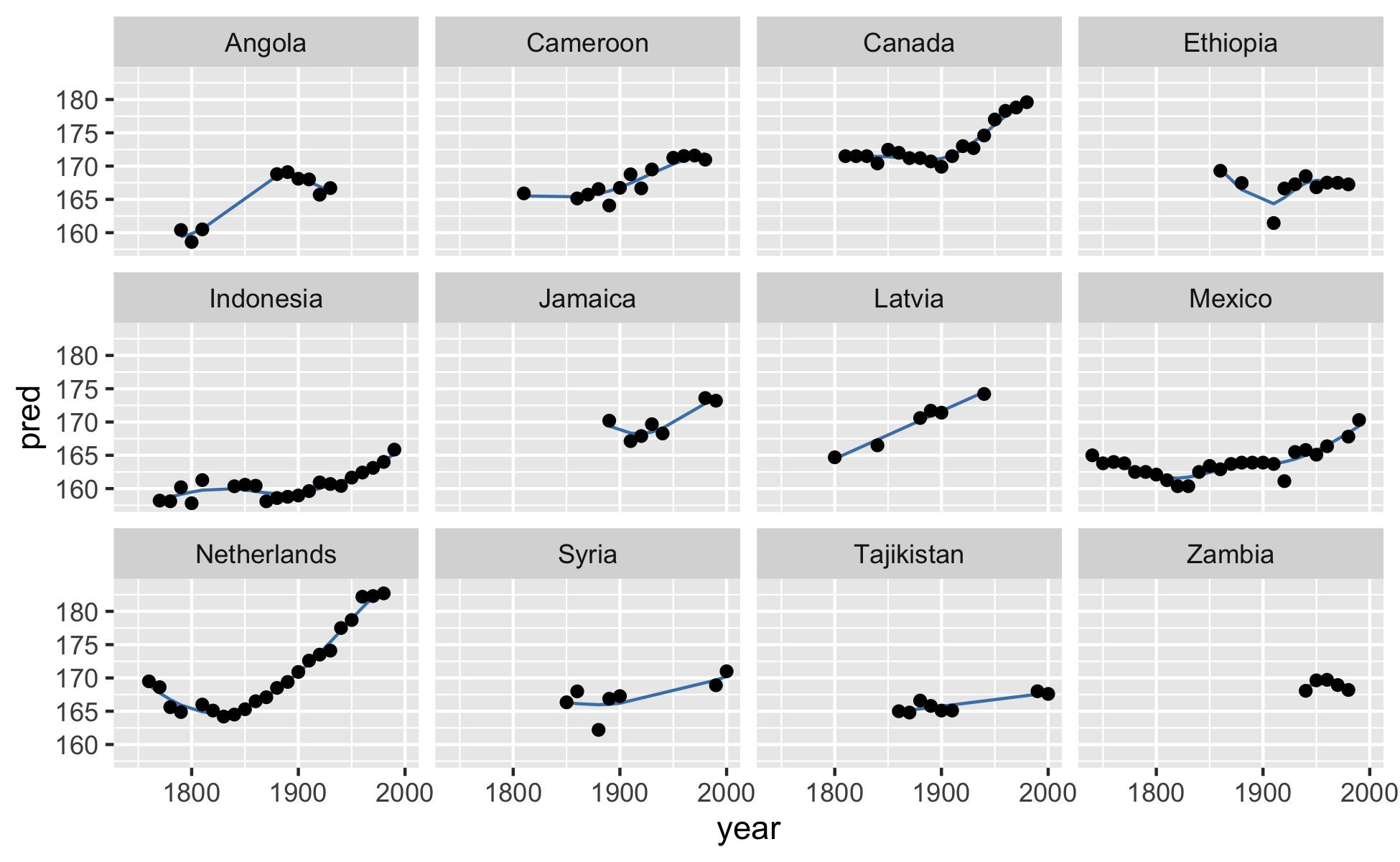} 

}

\caption[Exploration of a random sample of the data]{Exploration of a random sample of the data. This shows the data points of 12 countries, with the model fit in blue.}\label{fig:sample-n-keys-model}
\end{figure}
\end{Schunk}

\hypertarget{steretyping}{%
\section{Stereotyping}\label{steretyping}}

To help understand a population of measurements over time, it can be
useful to understand which individual measurements are typical (or
``stereotypical'') of a measurement. For example, to understand which
individuals are stereotypical of a statistic such as the minimum,
median, and maximum height. This section discusses how to find these
stereotypes in the data.

Figure \ref{fig:heights-aug-res-summary} shows the residuals of the
simple model fit to the data in the previous section. There is an
overlaid five number summary, showing the minimum, 1st quantile, median,
3rd quantile, and maximum residual value residuals, as well as a rug
plot to show the data. We can use these residuals to understand the
stereotypes of the data - those individuals in the model that best match
to this five number summary.

\begin{Schunk}
\begin{figure}

{\centering \includegraphics[width=0.6\linewidth]{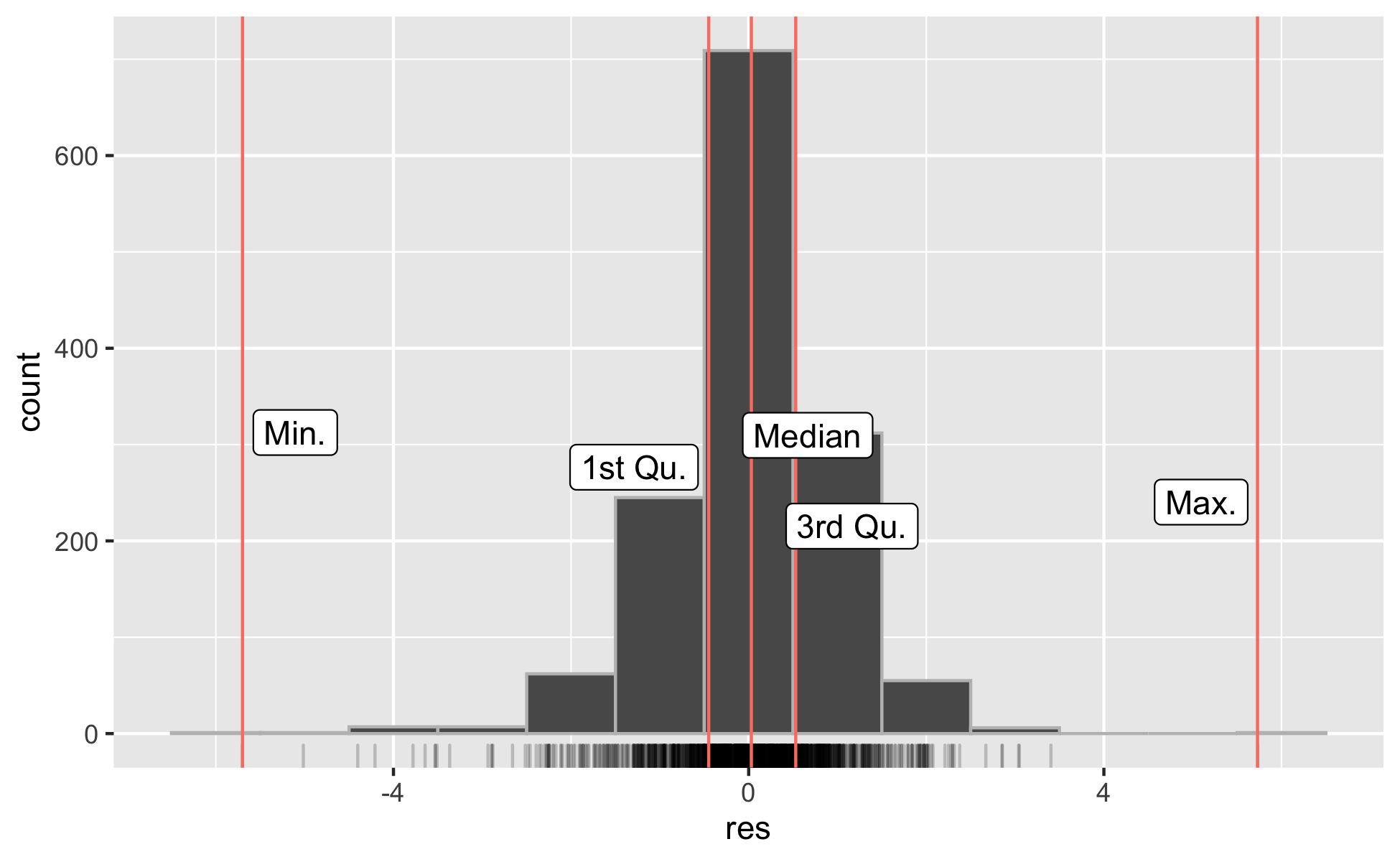} 

}

\caption[Five number summary of residual values from the model fit]{Five number summary of residual values from the model fit. The residuals are centered around zero with some variation.}\label{fig:heights-aug-res-summary}
\end{figure}
\end{Schunk}

We can do this using \texttt{keys\_near()} from \texttt{brolgar}. By
default this uses the 5 number summary, but any function can be used.
You specify the variable you want to find the keys nearest, in this case
\texttt{rss}, residual sums of squares for each key:

\begin{Schunk}
\begin{Sinput}
keys_near(heights_aug, var = rss)
\end{Sinput}
\begin{Soutput}
#> # A tibble: 62 x 5
#>    country   rss stat  stat_value stat_diff
#>    <chr>   <dbl> <fct>      <dbl>     <dbl>
#>  1 Denmark  9.54 med         9.54         0
#>  2 Denmark  9.54 med         9.54         0
#>  3 Denmark  9.54 med         9.54         0
#>  4 Denmark  9.54 med         9.54         0
#>  5 Denmark  9.54 med         9.54         0
#>  6 Denmark  9.54 med         9.54         0
#>  7 Denmark  9.54 med         9.54         0
#>  8 Denmark  9.54 med         9.54         0
#>  9 Denmark  9.54 med         9.54         0
#> 10 Denmark  9.54 med         9.54         0
#> # ... with 52 more rows
\end{Soutput}
\end{Schunk}

To plot the data, they need to be joined back to the original data, we
use a left join, joining by country.

\begin{Schunk}
\begin{Sinput}
heights_near_aug <- heights_aug %>% 
  keys_near(var = rss) %>% 
  left_join(heights_aug, 
            by = c("country"))
\end{Sinput}
\end{Schunk}

Figure \ref{fig:heights-keys-near} shows those countries closest to the
five number summary. Observing this, we see that the minimum RSS for
Moldova fits a nearly perfectly straight line, and the maximum residuals
for Myanmar have wide spread of values.

\begin{Schunk}
\begin{Sinput}
ggplot(heights_near_aug,
       aes(x = year,
           y = pred,
           group = country,
           colour = country)) + 
  geom_line(colour = "orange") + 
  geom_point(aes(y = height_cm)) + 
  scale_x_continuous(breaks = c(1780, 1880, 1980)) +
  facet_wrap(~stat + country,
             labeller = label_glue("Country: {country} \nNearest to \n{stat} RSS"),
             nrow = 1) + 
  theme(legend.position = "none",
        aspect.ratio = 1)
\end{Sinput}
\begin{figure}

{\centering \includegraphics[width=0.95\linewidth]{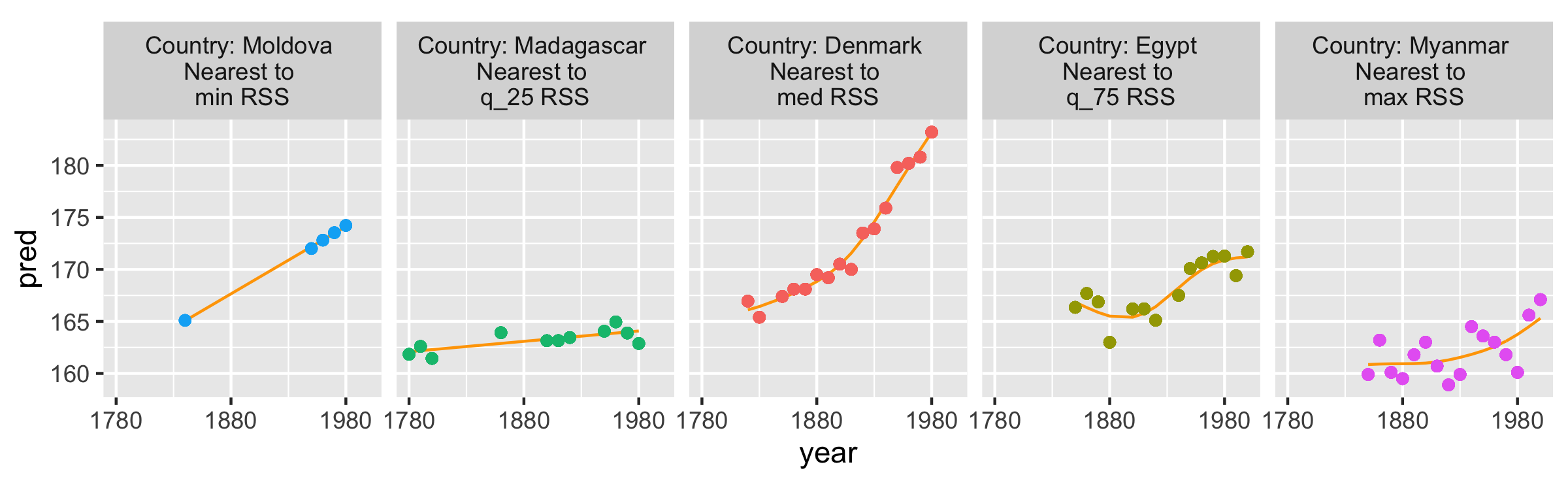} 

}

\caption[The keys nearest to the five number summary of the residual sums of squares]{The keys nearest to the five number summary of the residual sums of squares. Moldova and Madagascar are well fit by the model, and are fit by a straight line. The remaining countries with poorer fit have greater variation in height. It is not clear how a better model fit could be achieved.}\label{fig:heights-keys-near}
\end{figure}
\end{Schunk}

We can also look at the highest and lowest 3 residual sums of squares:

\begin{Schunk}
\begin{Sinput}
heights_near_aug_top_3 <- heights_aug %>% 
  distinct(country, rss) %>% 
  top_n(n = 3,
        wt = rss)

heights_near_aug_bottom_3 <- heights_aug %>% 
  distinct(country, rss) %>% 
  top_n(n = -3,
        wt = rss)

heights_near_top_bot_3 <- bind_rows(highest_3 = heights_near_aug_top_3,
                                    lowest_3 = heights_near_aug_bottom_3,
                                    .id = "rank") %>% 
  left_join(heights_aug,
            by = c("country", "rss"))
\end{Sinput}
\end{Schunk}

Figure \ref{fig:heights-keys-near-fancy-label} shows the same
information as the previous plot, but with the 3 representative
countries for each statistic. This gives us more data on what the
stereotypically ``good'' and ``poor'' fitting countries to this model.

\begin{Schunk}
\begin{figure}

{\centering \includegraphics[width=0.95\linewidth]{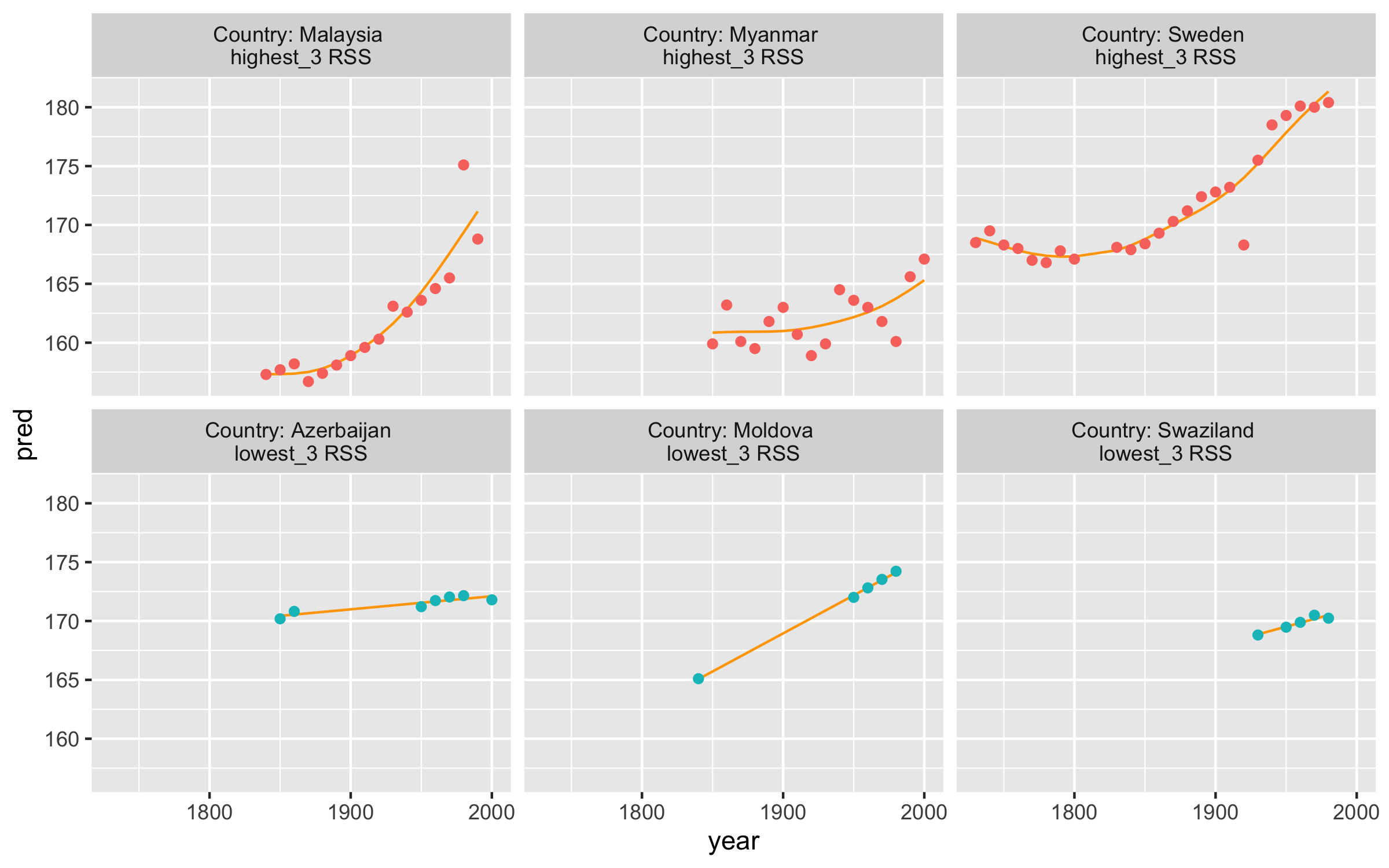} 

}

\caption[Figure of stereotypes for those keys with the three highest and lowest RSS values]{Figure of stereotypes for those keys with the three highest and lowest RSS values. Those that fit best tend to be linear, but those that fit worst have wider variation in heights.}\label{fig:heights-keys-near-fancy-label}
\end{figure}
\end{Schunk}

\hypertarget{install}{%
\section{Getting Started}\label{install}}

The \texttt{brolgar} R package can be installed from CRAN using

\begin{Schunk}
\begin{Sinput}
# From CRAN
install.packages("brolgar")
# Development version
remotes::install_github("njtierney/brolgar")
\end{Sinput}
\end{Schunk}

The functions are all designed to build upon existing packages, but are
predicated on working with \texttt{tsibble}. The package extends upon
\texttt{ggplot2} to provide facets for exploration:
\texttt{facet\_sample()} and \texttt{facet\_strata()}. Extending
\texttt{dplyr}'s \texttt{sample\_n()} and \texttt{sample\_frac()}
functions by providing sampling and stratifying based around keys:
\texttt{sample\_n\_keys()}, \texttt{sample\_frac\_keys()}, and
\texttt{stratify\_keys()}. New functions are focussed around the use of
\texttt{key}, for example \texttt{key\_slope()} to find the slope of
each key, and \texttt{keys\_near()} to find those keys near a summary
statistic. Finally, feature calculation is provided by building upon the
existing time series feature package, \texttt{feasts}.

To get started with \texttt{brolgar} you must first ensure your data is
specified as a \texttt{tsibble} - discussed earlier in the paper, there
is also a vignette
\href{http://brolgar.njtierney.com/articles/longitudinal-data-structures.html}{``Longitudinal
Data Structures''}, which discusses these ideas. The next step we
recommend is sampling some of your data with \texttt{facet\_sample()},
and \texttt{facet\_strata()}. When using \texttt{facet\_strata()},
facets can be arranged in order of a variable, using the \texttt{along}
argument, which can reveal interesting features.

To further explore longitudinal data, we recommend finding summary
features of each variable with \texttt{features}, and identifying
variables that are near summary statistics, using \texttt{keys\_near} to
find individuals stereotypical of a statistical value.

\hypertarget{summary}{%
\section{Summary}\label{summary}}

The \texttt{brolgar} package facilitates exploring longitudinal data in
R. It builds upon existing infrastructure from \texttt{tsibble}, and
\texttt{feasts}, which work within the \texttt{tidyverse} family of R
packages, as well as the newer, \texttt{tidyverts}, time series
packages. Users familiar with either of these package families will find
a lot of similarity in their use, and first time users will be able to
easily transition from \texttt{brolgar} to the \texttt{tidyverse} or
\texttt{tidyverts}.

Future work will explore more features and stratifications, and
stereotypes, and generalise the tools to work for data without time
components.

\hypertarget{acknowledgements}{%
\section*{Acknowledgements}\label{acknowledgements}}
\addcontentsline{toc}{section}{Acknowledgements}

We would like to thank Stuart Lee, Mitchell O'Hara Wild, Earo Wang, and
Miles McBain for their discussion on the design of \texttt{brolgar}. We
would also like to thank Rob Hyndman, Monash University and ACEMS for
their support of this research.

\hypertarget{paper-source}{%
\section*{Paper Source}\label{paper-source}}
\addcontentsline{toc}{section}{Paper Source}

The complete source files for the paper can be found at
\url{https://github.com/njtierney/rjournal-brolgar}. The paper is built
using rmarkdown, \texttt{targets} and \texttt{capsule} to ensure R
package versions are the same. See the README file on the github
repository for details on recreating the paper.

\bibliography{brolgar-paper.bib}

\address{%
Nicholas Tierney\\
Monash University\\%
Department of Econometrics and Business Statistics\\
ACEMS\\%
ACEMS Brisbane\\
\url{https://njtierney.com}%
\\\textit{ORCiD: \href{https://orcid.org/0000-0003-1460-8722}{0000-0003-1460-8722}}%
\\\href{mailto:nicholas.tierney@gmail.com}{\nolinkurl{nicholas.tierney@gmail.com}}
}

\address{%
Dianne Cook\\
Monash University\\%
Department of Econometrics and Business Statistics\\
ACEMS\\%
ACEMS Brisbane\\
\url{https://dicook.org}%
\\\textit{ORCiD: \href{https://orcid.org/0000-0002-3813-7155}{0000-0002-3813-7155}}%
\\\href{mailto:dicook@monash.edu}{\nolinkurl{dicook@monash.edu}}
}

\address{%
Tania Prvan\\
Macquarie University\\%
Department of Mathematics and Statistics\\
\\\textit{ORCiD: \href{https://orcid.org/0000-0002-6403-4344}{0000-0002-6403-4344}}%
\\\href{mailto:tania.prvan@mq.edu.au}{\nolinkurl{tania.prvan@mq.edu.au}}
}

\end{article}

\end{document}